\documentclass[10pt,twocolumn,floatfix,prb,superscriptaddress,aps]{revtex4-2}

\usepackage{physics}

\usepackage{color,newfloat}
\usepackage{graphicx}
\usepackage{dcolumn}
\usepackage{bm}
\usepackage{comment}
\usepackage{amsmath,amssymb,amsthm}
\usepackage{mathtools}
\usepackage{multirow}
\usepackage[dvipsnames]{xcolor}
\usepackage{dcolumn,url}

\usepackage{array}
\usepackage{makecell}
\usepackage{tabularx}

\usepackage[T1]{fontenc} 
\usepackage{txfonts}

\usepackage{braket}
\usepackage{physics}

\usepackage{hyperref}
\hypersetup{
    colorlinks=true,
    linkcolor=blue,
    citecolor=red,
    urlcolor=Maroon
}
\newcommand{\SM}{Supplementary information}

\newcommand{\Fbell}{90.03(4)}  
\newcommand{\FThree}{77.1(3)} 
\newcommand{\FFour}{66.5(9)} 
\newcommand{\Sraw}{2.33(1)}

\begin{document}

\title{Deterministic Multiphoton Entanglement Towards Fault Tolerance}
\newcommand{\AffCPH}{Center for Hybrid Quantum Networks (Hy-Q), The Niels Bohr Institute, University~of~Copenhagen,  DK-2100  Copenhagen~{\O}, Denmark}
\newcommand{\AffSQ}{Sparrow Quantum, Nordre Fasanvej 215, DK-2000 Frederiksberg, Denmark}
\newcommand{\AffBochum}{Experimental Physics VI, Ruhr-Universit\"at Bochum, Universit\"atsstra\ss e 150, 44801 Bochum, Germany}

\author{Bálint Sárközi$^{*}$}
\affiliation{\AffCPH{}}
\author{Ming Lai Chan}
\thanks{These authors contributed equally to this work.}
\affiliation{\AffSQ{}}
\author{Beatrice Polacchi}
\affiliation{\AffSQ{}}
\author{Jacob Esmann Poulsen}
\affiliation{\AffSQ{}}
\author{Matthias Christian Löbl}
\affiliation{\AffSQ{}}
\author{Zhe Liu}
\affiliation{\AffSQ{}}
\affiliation{\AffCPH{}}
\author{Martin Hayhurst Appel}
\affiliation{\AffCPH{}}
\author{Nikolai~Bart}
\affiliation{\AffBochum{}}
\author{Arne~Ludwig}
\affiliation{\AffBochum{}}

\author{Alexey Tiranov}
\affiliation{\AffSQ{}}
\author{Anders Søndberg Sørensen}
\affiliation{\AffCPH{}}
\author{Peter Lodahl}
\thanks{balint.sarkozi@nbi.ku.dk;\\ ming-lai.chan@sparrowquantum.com;\\peter.lodahl@sparrowquantum.com}
\affiliation{\AffCPH{}}
\affiliation{\AffSQ{}}

\date{}

\begin{abstract}

Entanglement is the quintessence of quantum science and technology, and photons are the natural carriers of quantum information enabling loophole-free Bell-inequality violation~\cite{Hensen2015}, photonic quantum simulators and computers~\cite{Alexander2025}, or networked quantum processors~\cite{Zhou2026}. 
The lack of deterministic high-fidelity multiphoton entanglement sources has been a major shortcoming limiting progress in photonic quantum technologies. An addressable spin confined in a single quantum emitter is the natural mediator of deterministic multiphoton entanglement~\cite{Gheri1998,Lindner2009}.  
Here, we realize a deterministic semiconductor spin-photon interface to generate high-fidelity multiphoton entanglement. The interface is based on a deterministically addressable single electron of a quantum dot embedded in a planar nanophotonic device. By integrating nuclear spin cooling and dynamical decoupling with engineered optical transitions, our device simultaneously combines near-perfect spin control with efficient emission of indistinguishable photons, enabling spin-photon entanglement with a fidelity above $90\%$ and a measured Bell-inequality violation.
We demonstrate deterministic generation of indistinguishable many-photon entangled states, directly validated up to $4$ qubits and projected entanglement across $12$ qubits. 
The in-depth residual infidelity and loss analysis identifies a clear and manufacturable pathway to fault-tolerant operation, leveraging a fundamentally new and scalable route to fault-tolerant quantum computing~\cite{Bartolucci2023} and quantum interconnects~\cite{Hartung2026}.

\end{abstract}

\date{\today}

\maketitle

\section{Introduction}

Photonics is particularly promising for fault-tolerant quantum computing due to its ability to realize long-range connections required for transversal logical gates~\cite{Eastin2009}, qLDPC codes~\cite{Bravyi2024, Chen2026}, and distributed quantum computing~\cite{Kimble2008}. Photonic qubits are decoherence-free and photonic processor technology is scalable and manufacturable, yet photon generation and two-qubit gate operations are intrinsically probabilistic leading to massive resource overhead when scaling up~\cite{Bartolucci2021switch,Lobl2026b}.  This picture changes radically with the advent of deterministic spin-photon interfaces: a single spin serves as a reusable quantum memory, sequentially entangling the emitted photons to generate multiphoton resource states on demand~\cite{Gheri1998,Lindner2009}.  Photon sources based on semiconductor quantum dots (QDs) embedded in photonic waveguides and cavities~\cite{Uppu2021} or trapped and laser-cooled single atoms in external cavities~\cite{Reiserer2015} enable deterministic photon generation, with high fibre-coupled system-efficiencies~\cite{Loredo2026}. QDs are particularly attractive platforms, combining ultrafast emission rates with manufacturability, scalability and compatibility with photonic integrated circuits~\cite{Uppu2021,Dewitte2025}, while colour centres~\cite{Nguyen2019} or single molecules~\cite{Roggors2026} represent alternative yet less mature solid-state photon emitters. Unlocking their full potential as resource-state generators, however, requires the simultaneous integration of high-fidelity spin control, indistinguishable photon emission, and high-quality entanglement generation, which so far has been elusive. 
   
Despite substantial advances, no solid-state spin-photon interface has yet combined these capabilities within a single device to realize deterministic multiphoton entanglement generation in a scalable manner.
In semiconductor QDs, coupling to the nuclear spin environment drastically reduces spin coherence, whereas unwanted optical decay channels may introduce errors during photon emission. One approach exploits polarization-encoded schemes enabling rapid photon generation~\cite{Schwartz2016,Huet2026,Laccotripes2024}, yet face an intrinsic trade-off since the required fast Zeeman-driven spin rotations degrade photon indistinguishability, thereby compromising the interference required for fusion gates~\cite{Bartolucci2023} underpinning actual applications. Furthermore, these schemes operate in a low magnetic field, which precludes effective nuclear spin cooling. On the other hand, the polarization approach excels in ease of operation and can be readily implemented in a planar waveguide geometry using chiral light-matter coupling \cite{stfeldt2022}.
An alternative approach based on time-bin encoding overcomes these trade-offs by operating on a single optical cycling transition utilizing, e.g., the intrinsic cyclicity naturally induced by the dipole-orientation-dependent Purcell effect in photonic-crystal waveguides (PCWs)~\cite{Tiurev2021}. The protocol combines nuclear spin cooling and spin-echo refocusing sequences to enable long spin coherence and high-fidelity spin rotations, which are prerequisites for multiphoton entanglement generation~\cite{Meng2024}.

\begin{figure*}
    \centering
    \includegraphics[width=\textwidth]{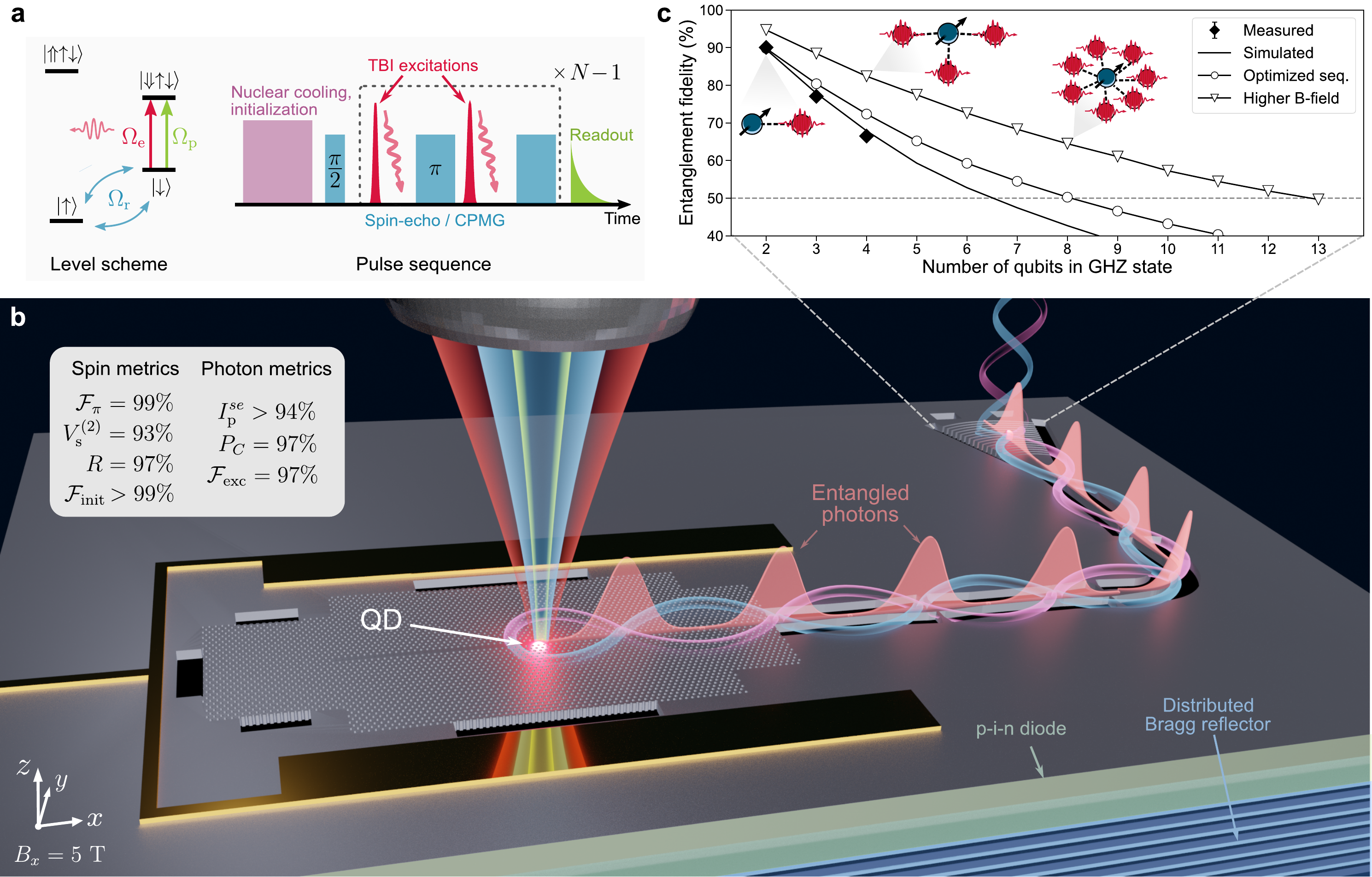}
    \caption{\textbf{Deterministic spin-photon interface and multiphoton entanglement generation.} The device consists of a QD electron spin qubit in a planar single-sided PCW. \textbf{a)} Energy level scheme of the quantum dot, with pulse sequence used to deterministically generate spin-multiphoton entangled states in the time-bin encoding protocol. Following nuclear-spin cooling, the sequence manipulates the electron spin state with Raman pulses (Rabi frequency $\Omega_\text{r}$) and employs optical excitation pulses ($\Omega_\text{e}$) to emit photons into well-defined time bins. Optical pumping ($\Omega_\text{p}$) enables readout of the spin state, while a time-bin interferometer (TBI) is used to create excitation pulses and read out the photonic state. \textbf{b)} Illustration of the entangled-photon source. Optical pulses are directed toward the chip through a high-NA objective inside a cryostat, which supplies an in-plane magnetic field of $B_x=5$~T. The source emits photons entangled with the hosted electron spin, which are collected into an optical fiber through a grating coupler.
    The QD is embedded within a $p$-$i$-$n$ diode for charge noise suppression and deterministic loading of an electron. A distributed Bragg reflector is grown underneath the diode for increased photon collection efficiency. Top left panel shows device specifications of the spin-photon interface relevant to entanglement generation, approaching unity. Spin metrics include spin-rotation fidelity $\mathcal{F}_\pi$, spin-echo visibility $V^{(2)}_{\text{s}}$, CPMG per-pulse visibility retention $R$, and spin initialization fidelity $\mathcal{F}_{\text{init}}$. Photon metrics consist of photon indistinguishability $I^{se}_{\text{p}}$, optical cycling probability $P_C$, and optical excitation fidelity $\mathcal{F}_{\text{exc}}$. \textbf{c)} Fidelity of spin-multiphoton GHZ states as a function of number of qubits, with the equivalent graph-state representations. Black diamonds are measured fidelities, in good agreement with simulation (black solid line).
    The measured fidelities are slightly reduced by implementing modified pulse sequence introduced to circumvent the finite detector deadtime (see main text). With optimized pulse sequences (white circles), GHZ-state fidelities could remain above $50\%$ for up to 8 qubits. Increasing the magnetic field to  $9$~T could lead to $12$ entangled qubits (white triangles). }
    
    \label{fig:interface}
\end{figure*}

Here, we realize a deterministic high-fidelity semiconductor spin-photon interface that simultaneously achieves optimized spin control, indistinguishable photon emission, and entanglement generation, all in a single nanophotonic device. 
Nuclear spin cooling and tailored dynamical decoupling yield high-fidelity spin rotations and refocusing visibilities, enabling high-fidelity entanglement generation, and the PCW realizes a highly selective optical transition and efficient emission of indistinguishable single photons into a single optical mode.
We show that this combination enables high-fidelity time-bin encoded spin-photon entanglement of $>90 \%$, and report the first violation of the Clauser-Horne-Shimony-Holt (CHSH) Bell inequality in a deterministic solid-state emitter-based source. We furthermore realize multipartite Greenberger-Horne-Zeilinger (GHZ) entanglement, explicitly validated up to 4 qubits. As illustrated in Fig. \ref{fig:interface}, the present operating conditions could be directly extended to generate 8 entangled qubits, and $12$-qubit entanglement can be achieved, e.g., by increasing the magnetic field.
A detailed analysis of residual imperfections (see below) outlines a concrete route to advancing the source into the regime of fault-tolerant operation. Our work advances deterministic solid-state emitters beyond proof-of-principle entanglement generation by establishing a high-fidelity spin-photon interface as a source of entangled photonic resource states directly applicable to fusion-based quantum computing~\cite{Chan2026} and scalable quantum interconnects~\cite{Beukers2024}.

\section{Operational principles of the deterministic spin-photon interface}

Realizing a deterministic source of high-fidelity multiphoton entanglement requires a coherent spin-photon interface that combines efficient photon emission with precise control of a stationary spin qubit. Our device consists of a single electron spin confined in a semiconductor InAs QD embedded in a PCW~\cite{lodahl2015}, as illustrated in Fig.~\ref{fig:interface}. The QD is integrated into a low-noise $p$-$i$-$n$ diode, enabling deterministic charging of an electron and electrical tuning of its optical transitions. The PCW strongly couples the QD emission into a guided optical mode while enhancing the selectivity of the cycling transition. The device operates at 4~K under an in-plane magnetic field of 5~T, which splits the electron spin states and enables coherent spin manipulation (Fig.~\ref{fig:interface}a). Optical pulses applied from above the device control the spin and excite the QD, while the emitted photons are collected through a grating coupler (Fig.~\ref{fig:interface}b).

The electron spin acts as a quantum memory that mediates entanglement between consecutively emitted photons. Starting from a spin superposition, a sequence of optical excitations and spin rotations entangles the spin state with the emission time of a photon, producing a time-bin encoded spin-photon entangled state~\cite{Tiurev2021}. Repeating this elementary operation sequentially entangles emitted photons and the spin state, with the structure of generated states determined by the spin-control sequence. The number of entangled photons is therefore determined by the number of coherent emission cycles, making the preservation of spin coherence and optical selectivity essential for scaling the entangled state.

Each emission cycle introduces several possible sources of error. Imperfect spin rotations and dephasing reduce the state coherence, while unwanted optical transitions reduce spin-photon correlations. Photon distinguishability degrades the interference essential for photonic fusion gates. Imperfections in any of these processes accumulate as the entangled state grows. The impact of these errors on the fidelity of an $N$-qubit GHZ state can be approximated as~\cite{Tiurev2021} 
\begin{align}
    \mathcal{F}^{(N)}_{\text{GHZ}} \approx \frac{1}{2}\bigg[V_{\text{s}}^{(N)}\cdot(I^{se}_{\text{p}})^{N-1} + \mathcal{P}^{(N)}_Q\bigg]-\frac{N-3/2}{2(C+1)}-\mathcal{E}^{(N)}_{\text{exc}}, \label{eq:fidelity}
\end{align}
where $N\geq2$. The spin-refocusing visibility $V_{\text{s}}^{(N)}$ captures the preservation of spin coherence throughout the entanglement generation, which contains an inherent Carr-Purcell-Meiboom-Gill (CPMG)-like dynamical decoupling sequence; $I_{\text{p}}^{se}$ is the single-emitter photon indistinguishability; and $\mathcal{P}^{(N)}_Q$ describes the GHZ-state fidelity of the population terms bounded by the spin-rotation quality factor $Q$ (\SM). The remaining terms originate from finite optical cyclicity $C$ and imperfect optical excitation errors $\mathcal{E}^{(N)}_{\text{exc}}$. Eq.~(\ref{eq:fidelity}) thus identifies the key performance metrics governing entanglement generation: spin-rotation fidelity, spin-refocusing visibility, photon indistinguishability, optical cyclicity, and excitation fidelity. This work combines several complementary techniques to optimize these properties within a single device, with all key metrics exceeding $90\%$ (Fig.~\ref{fig:interface}b): Nuclear spin cooling suppresses the Overhauser field fluctuations in the QD, thereby extending the electron-spin dephasing time $T_2^*$, and high-fidelity spin rotations with dynamical decoupling preserve coherence throughout sequential photon emission. Simultaneously, the engineered PCW provides a highly selective cycling transition and strong coupling to the guided mode, enabling coherent optical excitation, indistinguishable photon emission, and efficient photon extraction. 

\begin{figure*}
    \centering
    \includegraphics[width=\textwidth]{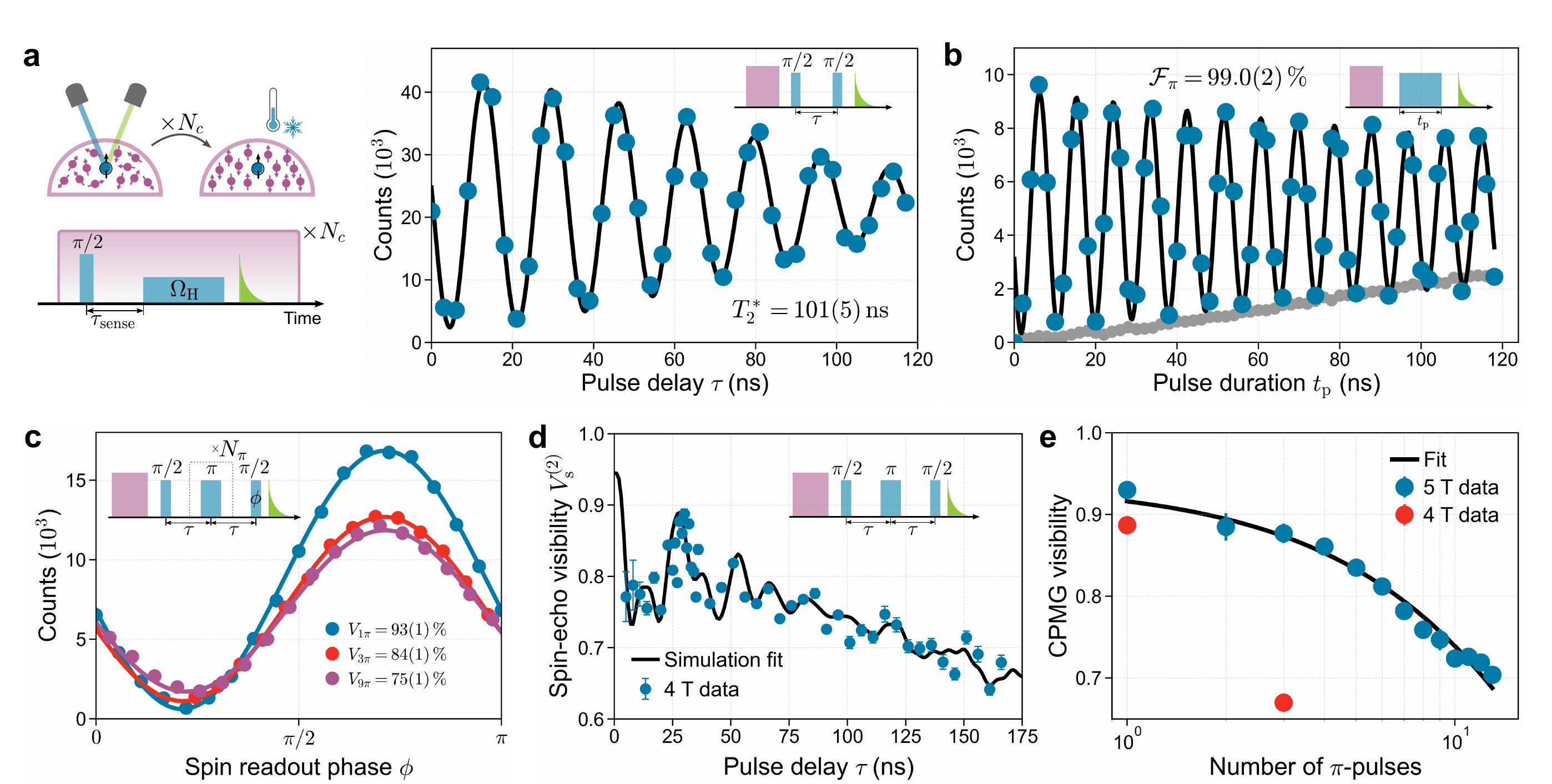}
    \caption{\textbf{Demonstration of optimized spin control.} 
    \textbf{a)} Ramsey spectroscopy performed after quantum-algorithmic cooling. Repeating the cooling pulse sequence (bottom left) $N_c$ times narrows noise fluctuations of the nuclear environment (top left), thus extending the electron dephasing time as shown in the Ramsey data (right). \textbf{b)} Spin Rabi oscillation performed with resonant (blue circles) and off-resonant (grey circles, $0.8$~GHz red-detuned)
    drives. The spin $\pi$-rotation fidelity is extracted from the fit which directly bounds $\mathcal{P}^{(N)}_Q$, the population contribution to the entanglement fidelity. 
    The off-resonant data closely follows the lower envelope of the Rabi oscillations, limited by laser-induced spin-flip processes. \textbf{c)} 
    Measured unnormalized CPMG visibility versus number of $\pi$-pulses $N_\pi$, which directly bounds the coherence contribution to the GHZ state fidelity as the number of qubits grows. Solid lines are sinusoidal fits used to extract the visibilities. For $N_\pi = 1$, the sequence reduces to a standard Hahn echo, with a record-high visibility of $V_{1\pi}=V^{(2)}_{\text{s}}=93(1)\%$.
    \textbf{d)} Spin-echo visibility as a function of pulse delay measured at $4$~T. Multiple visibility revivals arise at different delays due to relative Larmor precession of different nuclear spin species, the first of which is chosen as the pulse delay for entanglement sequences. 
    \textbf{e)} Measured peak CPMG visibility with increasing number of spin $\pi$-pulses at $4$~T and $5$~T, where the peak visibility decays significantly slower at a larger magnetic field strength. Data at 5~T is fitted with $V=V_0 R^{N_\pi}$ to extract a per-pulse visibility retention of $R=97.6(1)\%$.
    }
    \label{fig:Spin}
\end{figure*}

\section{High-fidelity spin control}
\label{sec:spin_control}

High-fidelity spin control is essential for maximizing the entangled-state fidelity through $\mathcal{P}^{(N)}_Q$ and $V^{(N)}_{\text{s}}$ in Eq.~(\ref{eq:fidelity}). We optimize both contributions by achieving near-unity spin-rotation fidelity and high refocusing visibilities for QD spins. 

To suppress nuclear noise that limits spin control, we first implement quantum-algorithmic cooling~\cite{Jackson2022}, in which the electron spin is used to sense and stabilize the collective state of the surrounding nuclei; see left panel of Fig.~\ref{fig:Spin}a. An electron is loaded into the QD, initialized by optical pumping, and prepared in a superposition via an optical Raman $\pi/2$-pulse to precess freely in the nuclear Overhauser field for a sensing time $\tau_\text{sense}$. A subsequent driving pulse at the Hartmann-Hahn resonance $\Omega_{\text{H}}$ enables an exchange interaction between the electron and the collective nuclear spin state~\cite{Bodey2019}. Resetting the electron spin after the exchange and repeating the sequence $N_c=100$ times cools the nuclear ensemble, narrowing fluctuations of the Overhauser field. A $57~\mu$s long cooling sequence extends the dephasing time $T_2^*$ from $\approx 2$~ns to $101(5)$~ns, measured using Ramsey spectroscopy of the electron spin (Fig.~\ref{fig:Spin}a). This is much longer than the duration $T_\pi=5.8$~ns of $\pi$-pulses used for entanglement generation, allowing for only $\mathcal{E}(T^*_2)=2(\pi T_2^*/T_\pi)^{-2}= 0.07$\% rotation error, thus enabling high-fidelity spin rotations.

We then estimate $\mathcal{P}^{(N)}_Q$ by measuring spin Rabi oscillations (Fig.~\ref{fig:Spin}b) and extract a quality factor of $Q=50(12)$, which describes the number of $\pi$-rotations within the exponential decay time of the oscillation envelope. This limits $\mathcal{P}^{(2)}_Q\leq97.6\%$, and bounds the $\pi$-rotation fidelity to $\mathcal{F}_{\pi}=(1+e^{-1/Q})/2 = 99.0(2)\%$~\cite{Bodey2019}, which exceeds previous reports for InAs QD spin~\cite{Hogg2025,Jackson2022,Meng2024} and is comparable to the performance in GaAs QD spin platforms~\cite{Appel2025}. In \SM~we include additional data to calibrate $Q$. The residual rotation infidelity ($1-\mathcal{F}_{\pi}-\mathcal{E}(T^*_2)=0.93\%$) is attributed to laser-induced spin-flip processes~\cite{Bodey2019} and optical Raman power instability. This is verified by repeating the Rabi measurement with an off-resonant drive red-detuned by 0.8 GHz (Fig.~\ref{fig:Spin}b, dark gray circles), wherein the off-resonant data tightly bounds the resonant decay envelope.

Beyond single-qubit rotations, the spin qubit needs to remain coherent over the period of entanglement generation. We employ CPMG sequences with $N_\pi$ refocusing $\pi$-pulses to filter out low-frequency nuclear spin noise and extend spin coherence beyond $T_2^*$. The resulting spin-refocusing visibility $V^{(N)}_{\text{s}}$ directly quantifies the spin coherence retained during entanglement generation (Eq.~\ref{eq:fidelity}). For $N_\pi=1$, the sequence reduces to a Hahn echo, yielding a record-high visibility of $V^{(2)}_{\mathrm{s}}=93(1)\%$ (Fig.~\ref{fig:Spin}c), which surpasses all previously measured values in QD platforms~\cite{Meng2024,Hogg2025,Stockill2016,Zaporski2023}. The high visibility is achieved by performing an iterative gradient-descent-like fine optimization of the Hahn-echo control pulse parameters and by increasing the magnetic field. We further show high visibilities $V^{(N)}_\text{s}$ with $N_\pi=3,9$, matching the sequences used to generate $N=3,4$-qubit GHZ states, respectively.

Scanning the Hahn-echo pulse delay at $B=4$~T reveals multiple revivals of the visibility (Fig.~\ref{fig:Spin}d), arising from the electron spin coupling to the coherently precessing nuclear spins through a strain-induced tilt between their quantization axes. This behaviour is well captured by our numerical model (black curve), which combines the calculated nuclear noise spectrum with the filter function of the applied pulse sequence~\cite{Stockill2016,Cywinski2008}, enabling quantitative predictions for different pulse sequences and magnetic fields. For entanglement generation, we operate at the first high-visibility revival, maximizing coherence retention and entanglement generation rate while preserving sufficient time to alternate spin rotations with photon emission. The key metric is therefore not the spin coherence time $T_2$ that characterizes long-time decay, but the unnormalized refocusing visibility at the shortest usable delay. Thus, despite the generally shorter $T_2$ of InAs QD spins compared with droplet-etched GaAs counterparts~\cite{Zaporski2023,Appel2025}, near-unity refocusing visibility---and therefore high entanglement fidelity---remains achievable.

Increasing the number of spin $\pi$-pulses in the CPMG sequence extends $T_2$ which allows more photons to be generated, but comes at the cost of reduced visibility due to accumulated pulse errors. We observe that the reduction heavily depends on the applied magnetic field strength. As seen in Fig.~\ref{fig:Spin}e, increasing the field strength from 4~T to 5~T leads to a significantly slower measured visibility decay, as verified by simulations (\SM). By fitting the measured data at 5~T with $V=V_0 R^{N_\pi}$, we extract a per-pulse visibility retention of $R=97.6(1)\%$, the same level reached in GaAs QDs~\cite{Zaporski2023}. This quantity ultimately bounds how $V^{(N)}_{\text{s}}$ scales with the number of qubits in the entangled state. We attribute the increase in $R$ with larger fields mainly to a more efficient filtering of nuclear noise, where the nuclear Zeeman interaction increasingly dominates over the strain-induced quadrupolar coupling and weakens the high-frequency Overhauser field components acting on the electron spin~\cite{Stockill2016}. Moreover, a wider spectral separation between the electron-spin resonance and nuclear sidebands suppresses unwanted off-resonant excitation within the spin-$\pi$ pulse bandwidth. In addition to these field-dependent effects, improved optical coupling raises $Q$ and further increases $R$ by reducing the infidelity from incoherent drive per pulse (\SM).

Overall, these results establish spin and optical performance metrics exceeding 90\% within a single device (Fig.~\ref{fig:interface}b). This includes spin-rotation fidelity of $\mathcal{F}_{\pi}=99.0(2)\%$, spin-echo visibility of $V^{(2)}_{\text{s}}=93(1)\%$, a CPMG per-pulse visibility retention of $R=97.6(1)\%$, spin initialization fidelity of $\mathcal{F}_{\text{init}}>99.2(1)\%$, raw Hong-Ou-Mandel visibility of $92.5(7)\%$, single-photon purity of $97.9(1)\%$, optical cycling probability of $P_C=\frac{C}{C+1}=97.4(1)\%$, and optical excitation fidelity of $1-\mathcal{E}^{(2)}_{\text{exc}}=97\%$. Additional measurements characterizing the above single-photon properties are presented in \SM, together with a comparison of these metrics against other state-of-the-art QD spin–photon interfaces. We next demonstrate how these combined capabilities translate into high-fidelity spin-photon and multiphoton entanglement generation.

\begin{figure*}
    \centering
    \includegraphics[width=\textwidth]{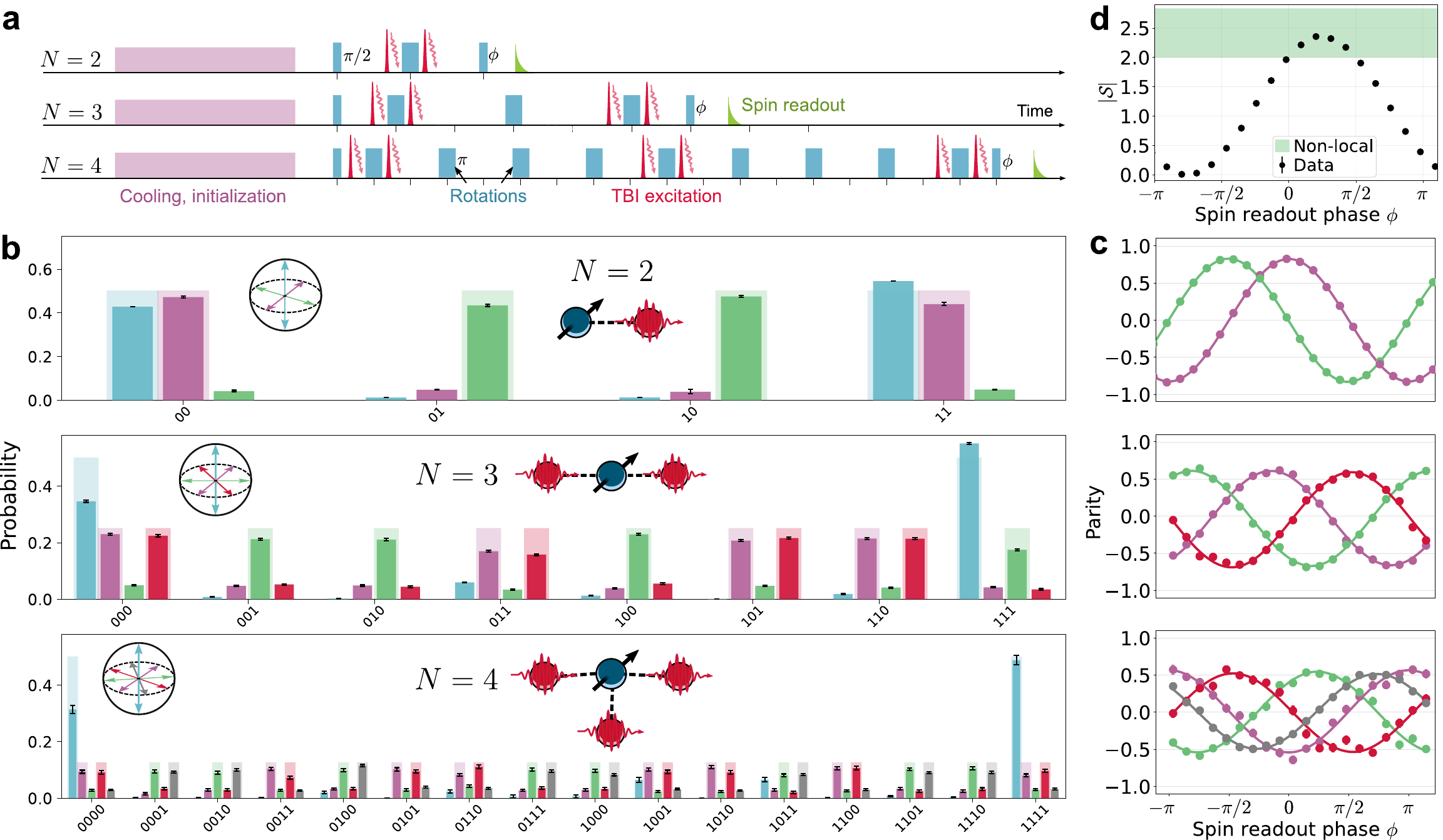}
    \caption{\textbf{Verification of spin-photon GHZ states. a)} Experimental pulse sequence producing $N$-qubit spin-photon GHZ states. Prior to state generation, a $57~\mu$s long nuclear-spin cooling sequence is first applied to both increase electron spin dephasing time and initialize its spin state. A series of spin rotation pulses interleaved with time-bin excitation pulses generates a Bell or GHZ state. The phase of the last spin $\pi/2$-pulse $\phi$ is swept before readout for spin projection on the equatorial basis. The duration of the full state generation sequence is set to $40~\mu$s, with around 40 state generation attempts per repetition. 
    To circumvent effects from finite detector deadtime, the pulse sequences for $N=3,4$ are modified. For $N=3$, the interpulse delay is extended to increase temporal separation between the two closest time bins. For $N=4$, two additional spin $\pi$-pulses are appended per photon. \textbf{b)} Experimental probability distributions of the joint spin-photon projective measurement outcomes for the generated GHZ state with varying $N$, compared with ideal values (shaded). Uncertainties are computed assuming Poissonian statistics. \textbf{c)} Experimental expectation values of the generated state measured in the equatorial basis, as a function of the spin readout phase $\phi$. Different colors correspond to different fixed photonic measurement bases. Circles are experimental data, while the solid lines are sinusoidal fits. Statistical uncertainties are plotted but not visible. \textbf{d)} CHSH Bell parameter $|S|$ as a function of $\phi$. The two photonic bases are set to $\sigma_z/\sigma_x$, while the two spin operators are $\cos(\phi_{0/1}) \sigma_x + \sin(\phi_{0/1}) \sigma_y$. Here $\phi=\phi_1$ is swept while $\phi_0$ is fixed to the optimal value.}
    \label{fig:Fig_3}
\end{figure*}

\section{High-fidelity spin-photon and multiphoton entanglement generation}
\label{sec:entanglement}

We generate and directly measure spin-multiphoton GHZ states of up to $N=4$ qubits, with results summarized in Fig.~\ref{fig:Fig_3}. 
The employed experimental pulse sequences for all values of $N$ are shown in Fig.~\ref{fig:Fig_3}a. A nuclear spin algorithmic cooling sequence is applied before state generation, which also initializes the spin in the ground state $\ket{\uparrow}$. Entanglement generation begins with a $\pi/2$-rotation preparing the spin in a superposition state $(\ket{\uparrow}+\ket{\downarrow})/\sqrt{2}$.
A picosecond laser pulse then drives the cycling transition between the ground state $\ket{\downarrow}$ and the excited trion state $\ket{\Downarrow\uparrow\downarrow}$,
resulting in photon emission in the early time bin $\ket{e}$ conditioned on $\ket{\downarrow}$. A subsequent $\pi$-rotation swaps the populations of $\ket{\uparrow}$ and $\ket{\downarrow}$, after which another picosecond pulse leads to conditional emission in the late time bin $\ket{l}$. This sequence is repeated a total of $N-1$ times, with an odd number of $\pi$-rotations in between repetitions. The rotation pulses are spaced to incorporate a CPMG sequence. The ideal GHZ state is therefore $\ket{\psi_{\text{GHZ}}} = (\ket{0}^{\bigotimes N} + \ket{1}^{\bigotimes N})/\sqrt{2}$ with $N-1$ photons and one spin qubit, where the spin state $\ket{\uparrow}$ ($\ket{\downarrow}$) and the photon state $\ket{e}$ ($\ket{l}$) encode the logical states $\ket{0}$ ($\ket{1}$), respectively.

The experimental GHZ state fidelity with the ideal state is estimated as
$\mathcal{F}^{(N)}_{\text{GHZ}} = (\mathcal{P}^{(N)} + \mathcal{C}^{(N)})/2$, with populations $\mathcal{P}^{(N)}$ and coherences $\mathcal{C}^{(N)}$ of the generated $N$-qubit density matrix determined by projecting each qubit in the respective measurement basis and recording $N$-photon coincidences (\SM). 
Intuitively, $\mathcal{C}^{(N)}$ is mostly affected by the first term in Eq.~(\ref{eq:fidelity}), while imperfections and asymmetry in $\mathcal{P}^{(N)}$ originate from optical excitation-induced spin-flip, cyclicity error and finite spin rotation $Q$-factor, as represented by the latter terms.

We report in Fig.~\ref{fig:Fig_3}b the complete set of $N$-qubit projective outcomes in the computational and equatorial bases for up to $N=4$. 
Fig.~\ref{fig:Fig_3}c shows the expectation value oscillations, or parity, of the measured equatorial operators  for $N=2,3,4$ by sweeping the relative phase $\phi$ of the last spin $\pi/2$-pulse before readout and modeled with a cosine function to estimate the coherences $\mathcal{C}^{(N)}$.
From here we obtain the following GHZ state fidelities:
$\mathcal{F}^{(2)}_{\text{Bell}}=\Fbell\%$, $\mathcal{F}^{(3)}_{\text{GHZ}}=\FThree\%$, and $\mathcal{F}^{(4)}_{\text{GHZ}}=\FFour\%$, 
and note that the values are not corrected for state preparation and measurement errors. Crucially, the measured fidelities are obtained with sub-optimal pulse sequences (Fig.~\ref{fig:Fig_3}a) to increase temporal separation between two consecutive photonic qubits to exceed the detector deadtime of $55$~ns. For $N=3$, a longer interpulse delay is chosen. For $N=4$, two additional $\pi$-rotations are included between excitation cycles at an optimal interpulse delay of $12.5$~ns. The deadtime constraint could be straightforwardly overcome by active routing of successive photons to alternating detectors. 
Using independently characterized spin and optical parameters (\SM), our microscopic model predicts that using optimally timed sequence without additional $\pi$-rotations, would enable fidelities exceeding the entanglement threshold of $50\%$ for up to 8 qubits and extending to $12$ by raising the magnetic field further (Fig.~\ref{fig:interface}c). This threshold can be relaxed by incorporating the full population distribution in Fig.~\ref{fig:Fig_3}b, enabling the verification of even larger entangled states~\cite{GuhneSeevinck2010}. We also note that higher fidelities are measured with optical excitation below $\pi$-pulse powers, where optical control errors are reduced at the cost of a lower excitation probability (\SM).

To further benchmark the quality of spin-photon entanglement generated by the source, we perform a Bell test by maximizing the CHSH inequality violation~\cite{Bell2004Speakable,Clauser1969} over the measured spin operator phase $\phi$, as shown in Fig.~\ref{fig:Fig_3}d. We find $\abs{\mathcal{S}} = \Sraw$, surpassing the local bound of 2 by $\approx33\sigma$. Our result constitutes the first Bell inequality violation from a spin-photon entangled state generated by a deterministic solid-state spin-photon interface, a prerequisite for scalable spin-multiphoton entanglement generation. In contrast, previous solid-state Bell tests, such as that reported in Ref.~\cite{Hensen2015}, relied on non-deterministic spin-photon interfaces.

For GHZ states with $N>2$, Mermin inequalities provide a fundamental test of multipartite quantum nonlocality \cite{Mermin1990}. A complete Mermin test is outside the scope of the present work. However, an estimate of the experimental Mermin parameter can be reconstructed from the measured three-qubit correlations, yielding $\mathcal{M}_{\mathrm{exp}}=2.572(2)$ which exceeds the local bound of 2~\cite{Mermin1990}. This suggests the potential for demonstrating multipartite nonlocality. Further details on experimental procedures for fidelity estimation, as well as Bell and Mermin tests are documented in \SM.  

\begin{figure*}
    \centering
    \includegraphics[width=\textwidth]{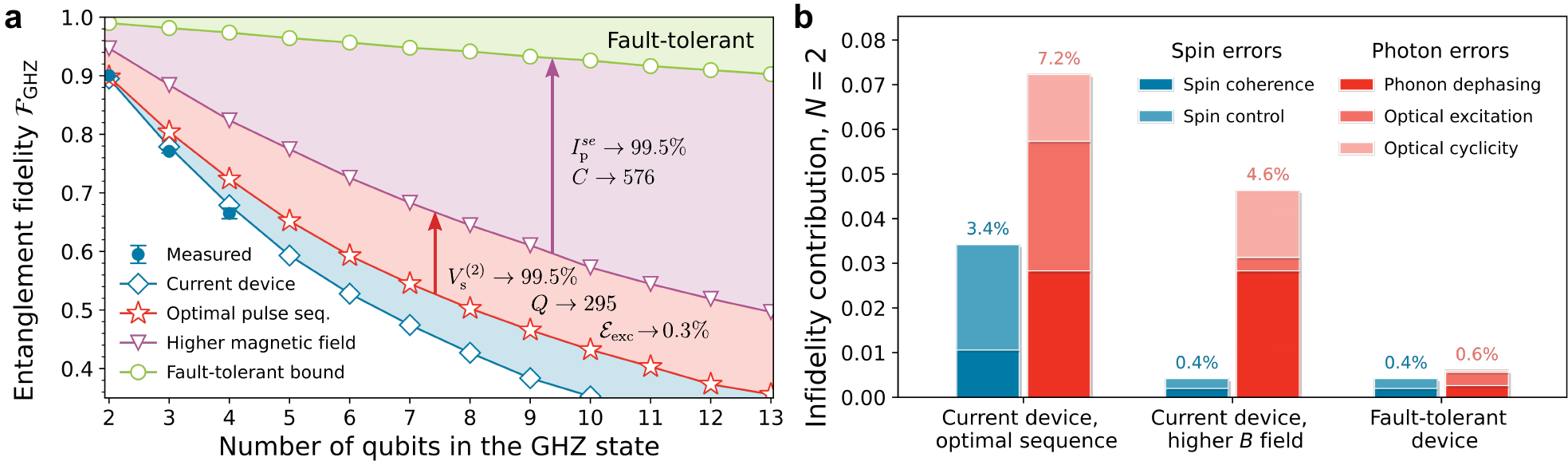}
    \caption{\textbf{Measured and projected multiphoton entanglement fidelities}.  \textbf{a)} GHZ state fidelity as a function of the number of qubits. Blue dots: measured data from the present device as it runs today. Blue diamonds: extension of the device’s performance to higher qubit numbers. Red stars: the current device with an optimally timed pulse sequence, free of detector dead time (cf. main text). Purple triangles: projection at a higher magnetic field of $9$~T along with the optimal pulse sequence, which simultaneously improves the spin properties and optical excitation fidelity with larger Zeeman splittings. Green circles: projection with higher cyclicity and indistinguishability, marking the start of the fault-tolerant regime. \textbf{b)} Simulated entanglement infidelity for Bell states, split into spin components (blue) and photonic components (red). Spin coherence error refers to contributions from imperfect echo rephasing, while spin control errors include spin-rotation and spin initialization infidelities. The spin contribution is expected to reach the fault-tolerant level with a larger magnetic field, while the photonic contribution requires both a higher cyclicity and further suppression of phonon-induced decoherence.
    }
    \label{fig:Projection_}
\end{figure*}

\section{Towards fault-tolerant operation}

We have reported high-fidelity spin-multiphoton entanglement generation and unraveled the physical mechanisms underlying each source of imperfection. We now discuss further advancements of the presented source to reduce these imperfections towards practical applications. The ultimate goal is the fault-tolerant regime allowing implementing quantum error correction to suppress physical errors faster than they accumulate. With deterministic spin-photon sources, resource-efficient and scalable roadmaps to photonic quantum computing~\cite{Chan2026} and one-way quantum repeaters~\cite{Borregaard2020} have been put forward. We consider two exemplary architectures, first, the architecture proposed in Ref.~\cite{Chan2026} where spins are used as data-qubits and, second, an all-photonic fusion-based architecture~\cite{Bartolucci2023}.

In Fig.~\ref{fig:Projection_}a, we compare the measured and simulated entanglement fidelities against the number of qubits. The agreement between experiment and the microscopic model indicates that the underlying physical processes are well captured, allowing device design and optimization. The error budget of Fig.~\ref{fig:Projection_}b shows that the Bell-state infidelity is dominated by photonic errors, namely phonon-induced pure dephasing ($2.8\%$), optical excitation errors ($2.9\%$) and finite optical cyclicity ($1.5\%$), whereas the spin contributions are significantly suppressed in this work, with imperfect echo rephasing ($1.1\%$), spin initialization and readout infidelities ($1.5\%$), and spin-rotation errors ($0.8\%$).
 
Monte Carlo simulations suggest that the spin-photon Bell-state fidelity can reach $\mathcal{F}^{(2)}_{\mathrm{Bell}}=95\%$ (Fig.~\ref{fig:Projection_}a, purple triangles) by suppressing optical errors and improving spin properties using established techniques. Applying a higher magnetic field (e.g., $9$~T) increases Zeeman splittings, thereby reducing the probability of exciting the other undesired cycling transition.  
Moreover, a higher field increases the spin coherence time due to the decrease of the nuclear spin noise amplitude and significantly improves the spin-echo rephasing efficiency~\cite{Stockill2016}.
Ultimately, strain inhomogeneity arising during QD growth limits the effectiveness of dynamical decoupling for InAs QDs \cite{chekhovich_suppression_2015}, which can be heavily suppressed by engineering low-strain QDs, allowing spin coherence time up to $0.1$~ms~\cite{Zaporski2023}. Further improvement in the spin quality factor towards $Q=295$ lies in reducing the incoherent component of the optical Raman drive. An optimized nanophotonic design with leaky modes at the Raman laser frequency could enhance coupling with the optical driving field, expected to significantly reduce the required amount of optical power thus improving the spin rotation fidelity. Together with a lower noise in microwave pulses, pulse shaping~\cite{Motzoi2009}, and engineered charge environment in the diode heterostructure, we estimate that $\mathcal{P}^{(2)}_Q>99.83\%$ is possible (Fig.~\ref{fig:Projection_}b). 

On the photonics side, the optical cyclicity can be further improved by tuning the magnetic field direction~\cite{Koong2026} and proper positioning of the QD in the PCW, towards the fault-tolerant threshold of $C=576$~\cite{Chan2026}. Furthermore, the minor phonon decoherence processes can be further reduced where $I^{se}_{\mathrm{p}}\to99.5\%$ is reachable by implementing vibrational clamping of PCWs and/or lowering the device operating temperature~\cite{Tighineanu2018}. Finally, near-unity source efficiency needs to be achieved to meet the stringent requirements of fault-tolerant quantum computing architectures~\cite{Chan2026}.  In addition to near-unity source fidelity, fault-tolerant operation requires photon loss to be reduced below an architecture-specific threshold. Furthermore, near-unity intrinsic coupling efficiency are readily achieved in  PCWs~\cite{Arcari2014} and very significant progress on ultra-low-loss transporting and outcoupling have been reported~\cite{Loredo2026} paving the route for fault-tolerance performance. More details on the above strategies and fidelity simulations are given in \SM.

In photonic quantum computing, a high flux source of high-fidelity entangled multiphoton states could replace a large network of multiplexed probabilistic sources, and entanglement purification~\cite{Lobl2026a} could further suppress error rates of the multiphoton states (\SM). As an example, we consider the 6-ring resource state proposed as the backbone for fusion-based quantum computing architectures~\cite{Bartolucci2023}. We estimate that a single deterministic source could produce encoded (unencoded) 6-ring states with a generation rate equivalent to about $97~(480)$ probabilistic down-conversion sources (\SM), assuming an interpulse delay of $\tau=12.5$~ns used in the current device. Notably this gap between deterministic and probabilistic sources will generally widen even further for generating more complex multi-photon entangled seed states. Looking ahead, the deterministic sources could be operated faster, and a delay of $\tau=1.2$~ns could retain near-unity CPMG visibility in a large magnetic field and using droplet-etched GaAs QDs~(\SM). The reduction in $\tau$ by almost an order of magnitude would amplify the overhead advantage by the same factor. Finally, the present work has focused on a single QD, but the planar waveguide-based platform opens the opportunity to scale to multiple QDs to realize more complex and powerful encoded resource states~\cite{Pettersson2025a}. These states could be generated through either by probabilistic fusions~\cite{Lee2023} or even deterministically with QD-QD gates~\cite{Li2022} potentially mediated directly by dipole-dipole interaction through the waveguide~\cite{Tiranov2023}. 

\section{Concluding remarks}

We have realized a high-fidelity spin-multiphoton interface based on a single semiconductor quantum dot embedded in a planar photonic-crystal waveguide. The simultaneous integration of optimized spin control with engineered light-matter interactions enables all key spin-photon metrics to exceed $90\%$ in a single device, defining a concrete standard for spin-multiphoton interfaces aimed at deterministic entanglement generation. These advancements translate directly into high-fidelity spin-photon entanglement sufficient for Bell-inequality violation and genuine multipartite entanglement explicitly validated to 4 qubits. Together with photon collection efficiencies of $>55\%$ already demonstrated in this platform~\cite{Loredo2026} and concrete improvement strategies identified, these results establish a promising route towards even larger detectable entangled states reaching fault-tolerance. 

Taking a broader perspective, when compared with various quantum-emitter platforms, including neutral atoms in optical cavities~\cite{Thomas2022}, color centers in diamond~\cite{Nguyen2019} or single molecules~\cite{Roggors2026}, the primary limitation for semiconductor quantum dots has traditionally been the comparatively short spin coherence, despite their ability to generate photons rapidly, coherently and efficiently into nanophotonic structures. Our results show that significant advances in spin control now allow these strengths to be combined: high spin-refocusing visibility is preserved over the short operating timescales relevant to photon generation despite the relatively modest $T_2$, enabling high entanglement fidelity alongside rapid emission of indistinguishable photons. 
This advantage is further complemented by the maturity of III--V nanofabrication~\cite{lodahl2015} and by heterogeneous integration of planar sources with photonic circuitry~\cite{Dewitte2025}, which provides a scalable route to deployment. 

\section*{Author Contributions}
B.S and M.L.C contributed equally to this work. B.S, M.L.C, and B.P built the experimental setup, performed the experiment and analyzed the data. J.P built the pulse stretcher and pulse picking setup. M.C.L performed resource estimations and simulations for purification protocols. Z.L fabricated the sample. M.H.A developed the experimental server architecture and contributed to the implementation of quantum-algorithmic cooling. A.L designed the heterostructure and directed its MBE growth. N.B grew the quantum dot wafer. A.T performed fidelity simulations. A.T, A.S.S and P.L supervised the project. B.S, M.L.C, B.P, M.C.L, A.T and P.L wrote the manuscript with inputs from all authors.

\section*{Acknowledgements} 
We gratefully acknowledge financial support from Danmarks Grundforskningsfond (DNRF 139, Hy-Q Center for Hybrid Quantum Networks), the Novo Nordisk Foundation (Challenge project ``Solid-Q”), and Danmarks Innovationsfond (IFD1003402609, FTQP). We thank Larissa Vertchenko for the Blender image in Fig.~1b. M.L.C acknowledges funding from Danmarks Innovationsfond (Grant No.~4298-00011B). M.L.C, B.P, and A.T acknowledge funding from HORIZON-EIC-ACC (101254310, QPAIR).

\clearpage

%

\end{document}